\documentclass[prl,aps,twocolumn]{revtex4}
\usepackage{amssymb}
\usepackage{graphicx}

\begin{document}

\title{Angle-dependent magnetoresistance 
in the weakly incoherent interlayer transport regime}
\author{M. V.~Kartsovnik$^1$}
\author{D.~Andres$^1$}
\author{S. V.~Simonov$^2$}
\author{W.~Biberacher$^1$}
\author{I.~Sheikin$^3$}
\author{N. D.~Kushch$^4$}
\author{H.~M\"uller$^5$}

\affiliation{$^1$Walther-Meissner-Institut, Bayerische Akademie der
Wissenschaften, D-85748 Garching, Germany}
\affiliation{$^2$ Institute of Solid State Physics, Russian Academy of 
Sciences, 142432 Chernogolovka, Russia}
\affiliation{$^3$ Grenoble High Magnetic Field Laboratory, CNRS,
BP 166, 38042 Grenoble Cedex 9, France}
\affiliation{$^4$ Institute of Problems of Chemical Physics, 
Russian Academy of Sciences, 142432 Chernoglovka, Russia}
\affiliation{$^5$European Synchrotron Radiation Facility, F-38043
Grenoble, France}

\begin{abstract}
We present comparative studies of the orientation effect of a 
strong magnetic field on the interlayer resistance of 
$\alpha$-(BEDT-TTF)$_2$KHg(SCN)$_4$ samples characterized by 
different crystal quality. We find striking differences in their 
behavior which is attributed to the breakdown of the coherent 
charge transport across the layers in the lower quality sample. 
In the latter case, the nonoscillating magnetoresistance 
background is essentially a function of only the out-of-plane 
field component, in contradiction to the existing theory.
\end{abstract}
\maketitle

The extremely high electronic anisotropy is a common feature of 
many exotic conductors extensively investigated in the recent 
years, such as, for example, organic conductors \cite{ishi98,wosn96} 
or layered metal oxide superconductors \cite{coop94,berg03,jin03}. 
The mechanism of 
the interlayer charge transfer is one of the central questions 
in understanding the nature of various ground states and 
electronic properties of these materials. 
In particular, the problem of discriminating between coherent 
and incoherent interlayer transport has received much attention 
(see, e.g., 
\cite{valla02a,kenz98a,osad02a,wosn02,sing02,kura03,huss03}). 

If the coupling is strong enough, so that the interlayer hopping 
time, $\tau_h \sim \hbar/t_{\perp}$, where $t_{\perp}$ is the 
interlayer transfer integral, is considerably shorter than the 
transport scattering time $\tau$, 
the electron transport is fully coherent and can be adequately 
described within the anisotropic three-dimensional (3D) Fermi 
liquid model. 
In the other limit, $\hbar/t_{\perp}\gg\tau$, the successive 
interlayer hopping events are uncorrelated; thus the electron 
momentum and the Fermi surface can only be defined in the 
plane of the layers. Here one should distinguish between two 
different transport regimes. In the {\em strongly} incoherent 
regime there is no interference between the electron wave 
functions on adjacent layers and the interlayer hopping is 
entirely caused by scattering processes. 
Consequently, the temperature dependent resistivity  
across the layers, $\rho_{\perp}(T)$, is nonmetallic. 
On the other hand, one can 
consider the case of a weak overlap of the wave functions 
on adjacent layers, so that the interlayer transport is mostly 
determined by one particle tunneling. This {\em weakly} 
incoherent transport was studied in a number of theoretical 
works \cite{kenz98a,osad02a,soda77,kuma92} assuming that the 
intralayer momentum is 
conserved during a single tunneling but successive tunneling 
events are uncorrelated due to scattering within the 
layers. The transverse resistivity $\rho_{\perp}$ has been 
shown to be almost identical to that in the coherent 
case, sharing with the latter the metallic temperature 
dependence \cite{soda77,kuma92} and most of high-field magnetotransport 
phenomena \cite{kenz98a,osad02a}. Thus the question arises: is there 
a substantial physical difference between the coherent and weakly 
incoherent interlayer transport regimes? 

Moses and McKenzie \cite{kenz98a} proposed to use  
the angle-dependent magnetoresistance to distinguish 
between the two cases: When the field 
is turned in a plane normal to the layers, a narrow peak is 
often observed at the orientations nearly parallel to the layers 
\cite{kart04a}. 
This so-called {\em coherence peak} is associated with a 
topological change of electron cyclotron orbits on a 3D Fermi 
surface slightly warped in the direction perpendicular to the 
layers \cite{kart04a,hana98,pesc99a} and can only exist in 
the coherent regime. Its absence in the weakly incoherent 
transport model \cite{kenz98a} is a natural consequence of the 
assumed strictly 2D Fermi surface. 

The observation of the coherence peak has been used as an  
argument for the coherent interlayer coupling in a number of 
layered conductors \cite{wosn02,sing02,huss03,kura03}. However, 
no systematic experimental study of the weakly incoherent 
regime has been done thus far. 
Here we present comparative studies of the 
orientation effect of a high magnetic field on the 
interlayer magnetoresistance of different samples of the 
layered organic conductor $\alpha$-(BEDT-TTF)$_2$KHg(SCN)$_4$. 
We argue that, depending on the crystal quality, either the 
coherent or weakly incoherent transport regime can be realized 
in this material. In agreement with the theoretical predictions, 
the coherence peak is only observed in the highest quality 
samples. 
However, by contrast to the coherent case, an important new 
feature, that cannot be explained by existing theories, 
has been found in the weakly incoherent regime: 
the magnetoresistance in a field strongly tilted towards the 
layers turns out to be insensitive to the inplane 
field component.

$\alpha$-(BEDT-TTF)$_2$KHg(SCN)$_4$ is one of the most 
anisotropic organic conductors \cite{kart04a}. Its electronic 
system comprises a quasi-1D and a quasi-2D conduction 
band. 
This compound exhibits a complex 
``magnetic field--pressure--temperature'' phase diagram 
which can be consistently explained 
by a charge-density-wave (CDW) formation at an unusually low 
temperature, $T_{\mathrm{CDW}}\approx 8$~K 
\cite{kenz97,chri00,andr01}. 
In the CDW state the quasi-1D carriers are gapped whereas  
the quasi-2D band remains metallic. Since we are presently 
focusing on the metallic magnetotransport, numerous 
anomalies associated with field-induced CDW transitions should 
be avoided. We will, therefore, consider the zero-pressure CDW 
state only at relatively low fields, up to 10~T, at which no 
significant change of the electronic system occurs. In addition, 
we present data taken at a high pressure, $P=6.2$~kbar, which 
suppresses the CDW, restoring the fully normal metallic  
state. The measurements have been done at $T=1.4$~K. 

Figure 1 illustrates the dependence of the interlayer 
resistance of two different samples on the magnetic field 
orientation, measured at zero pressure. 
The orientation is defined by the 
polar angle $\theta$ between the field direction and the normal 
to the plane of the layers, and by the azimuthal angle $\varphi$ 
between the projection of the field on the plane and the 
crystallographic $a$ axis. Both samples exhibit prominent 
angular magnetoresistance oscillations (AMROs) periodic 
in the $\tan\theta$ scale: the resistance sharply drops at 
the Lebed magic angles \cite{lebe86a}. This behavior 
is well known for the present material and is associated with the 
open-orbit motion of the metallic quasi-2D carriers in the presence 
of a CDW potential \cite{kart96}. What we 
want to focus on now is the nonoscillating 
background which turns out to be drastically different in the 
two samples shown in Fig.~1. 
\begin{figure}[t]
\includegraphics[width=.85\linewidth]{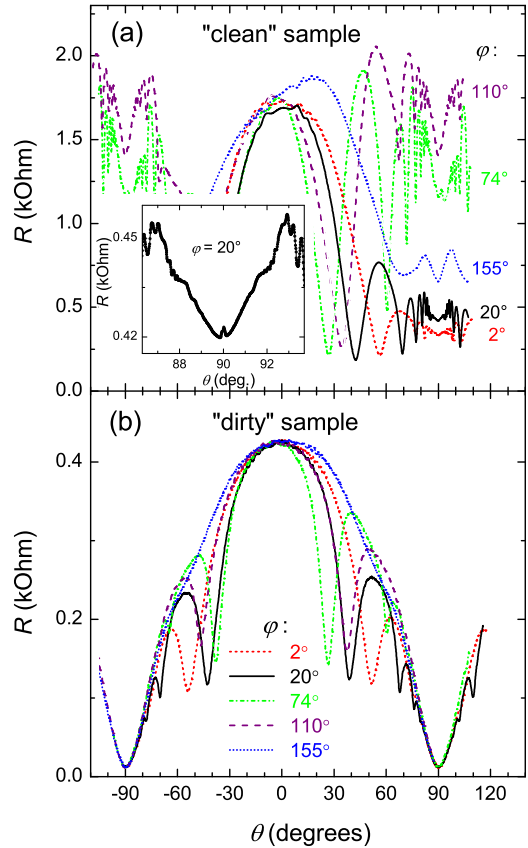}
\caption{(color online). Interlayer resistance of two crystals 
of $\alpha$-(BEDT-TTF)$_2$KHg(SCN)$_4$ as a function of the 
polar angle $\theta$ recorded at different azimuthal 
angles $\varphi$, $P=0$~kbar, $B=10$~T. Inset in the 
upper panel: details of a $\theta$-sweep for the "clean" 
sample, showing the small coherence peak.}
\label{fig1}
\end{figure} 

In the highest quality sample (see Fig.~1a) the nonoscillating 
magnetoresistance component displays a rather complex behavior 
strongly depending on the azimuthal angle $\varphi$. In particular, 
the $\varphi$-dependence of the resistance at the field aligned 
exactly parallel to the layers, i.e. at $\theta = 90^{\circ}$, 
is directly related to the in-plane curvature of the Fermi surface 
\cite{lebe97,kart04a}. Further, a detailed inspection of the 
$\theta$-dependence around $\theta = 90^{\circ}$ reveals a very 
narrow peak as shown in the inset in Fig.~1a. It is, 
to our knowledge, the first observation of the coherence peak in 
the present compound. The peak has been found at the azimuthal 
orientations, $0^{\circ}\leq \varphi \leq 50^{\circ}$, its width 
$\Delta \theta$ varying between $0.12^{\circ}$ and $0.35^{\circ}$. 
One can, therefore, estimate the Fermi surface corrugation   
in the interlayer direction \cite{kart04a}: 
$\Delta k_{\|}/k_F \approx \Delta\theta/k_Fd 
\simeq 1.5\times 10^{-3}$, 
where we have taken the mean value $\Delta \theta = 0.23^{\circ}$, 
the intralayer Fermi wave number $k_F\simeq 0.14$~\AA$^{-1}$, 
and the interlayer spacing $d\approx 20$~\AA\ \cite{wosn96}. 
Further, estimating roughly the Fermi energy 
from the de Haas--van Alphen data \cite{wosn96}, 
$\varepsilon_F \sim 40$~meV, we arrive at an extremely 
low value of the interlayer transfer integral: 
$t_{\perp} \approx (\Delta k_{\|}/2k_F)\varepsilon_F 
\sim 0.03$~meV. 

Another kind of the angular dependence is observed on 
the second sample as illustrated in Fig.~1b. The amplitude 
of the AMRO is considerably weaker here and the oscillations 
are damped, with tilting the field towards $\pm 90^{\circ}$, 
much faster than in the previous case, thus indicating a lower 
crystal quality. We, therefore, will refer to this sample as to 
the "dirty" one, by contrast to the "clean" sample considered 
above \cite{comment1}. Note, however, that both samples are 
clean enough in the sense that the strong field criterion, 
$\omega_c\tau\gg 1$, is always fulfilled in fields of a few tesla. 

No coherence peak has been found for the "dirty" sample at any 
$\varphi$. According to the theory \cite{kenz98a,osad02a}, 
this means the breakdown of the interlayer coherence. 
On the other hand, the presence of AMRO and the metallic 
temperature dependence $R(T)$ indicate the {\em weakly} 
rather than {\em strongly} incoherent interlayer transport 
regime to be realized in the present case. 

The most obvious distinction of the "dirty" sample is the 
behavior of the nonoscillating magnetoresistance background: 
the latter decreases steadily as the field is tilted towards 
the layers, producing a broad dip around 
$\theta = \pm 90^{\circ}$. Remarkably, as seen from Fig.~1b, 
this behavior is practically independent of the azimuthal 
orientation of the field rotation plane.

The results above were obtained at zero pressure, 
in the partially metallic CDW state. To verify that the drastic  
difference in the behavior of the "clean" and "dirty" samples is 
related to the metallic magnetotransport and not to some specific 
features of the CDW state, we have performed measurements under 
high pressure at which the whole material is entirely 
normal metallic. 

\begin{figure}[t]
\includegraphics[width=.85\linewidth]{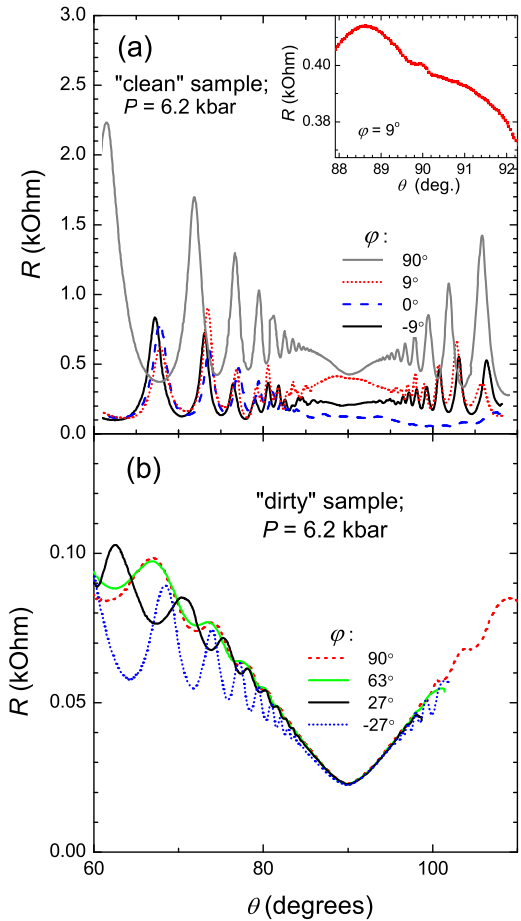}
\caption{(color online). Angle-dependent magnetoresistance of 
(a) the "clean" sample and (b) the "dirty" sample at 
$P=6.2$~kbar, $B=20$~T. Inset in the upper panel: fragment 
of the $\varphi = 9^{\circ}$ curve for the "clean" sample 
with the coherence peak.}
\label{fig2}
\end{figure}
Examples of the $\theta$-sweeps recorded for "clean" and "dirty" 
samples at the pressure of 6.2~kbar are shown in 
Fig. 2. The Fermi surface and, therefore, the electron 
orbit topology are different from those at zero pressure. This is, 
in particular, reflected in the AMRO behavior \cite{kart95d,hana96}: 
now the oscillations 
are mostly determined by closed orbits on the cylindrical Fermi 
surface. Despite the radical 
modification of the magnetoresistance behavior upon applying 
pressure, the major differences between the "clean" and "dirty" 
samples remain the same as in the zero-pressure state. The 
"clean" sample exhibits a small narrow peak around 
$\theta = 90^{\circ}$ (see the inset in Fig. 2,a) and a strong 
dependence on the azimuthal orientation $\varphi$. By contrast, 
the "dirty" sample shows no coherence peak and is insensitive 
to $\varphi$ at sufficiently high tilt angles 
$\theta$. 

\begin{figure}[t]
\includegraphics[width=.85\linewidth]{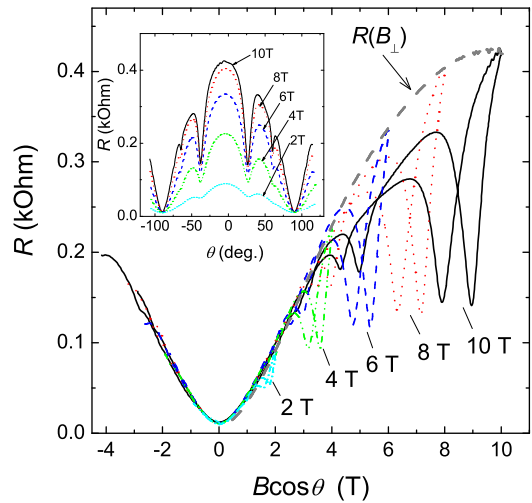}
\caption{(color online). Magnetoresistance of the "dirty" 
sample at $P=0$ as a function of the out-of-plane field 
component. The raw $\theta$-sweeps recorded at different 
field strengths are shown in the inset.}
\label{fig3}
\end{figure}
The decrease of the magnetoresistance of the "dirty" sample, as 
the field direction approaches the plane of the layers, and its 
independence of the azimuthal angle $\varphi$ suggests 
that it does not feel the magnetic field component parallel to 
the layers. 
To check this, we have made $\theta$-sweeps at different 
values of the field strength and replotted the resistance as 
a function of the field projection on the normal to the layers, 
$B_{\perp} = B\cos\theta$. The result for zero pressure is shown 
in Fig.~3. 
Except the vicinities of the magic angles, all the curves, recorded 
at fields from 2 to 10~T, collapse on a single line. 
A similar behavior is observed at higher fields
in the high-pressure state. Moreover, the curves shown in Fig. 3 
nicely coincide with the field dependence $R(B)$ taken at 
the field perpendicular to the layers (dashed gray line in Fig. 3). 
Thus, we conclude that the magnetoresistance of the "dirty" sample 
at high tilt angles is essentially a function of only the 
field component perpendicular to the layers. 

This is a surprising and somehow counterintuitive result. Normally, 
an in-plane magnetic field acts to confine charge carriers to 
the layers, thus increasing the interlayer resistivity. The 
theory predicts a strong linear or superlinear magnetoresistance 
in strong fields parallel to the layers, both in the coherent 
\cite{lebe97,pesc97} and weakly incoherent \cite{kenz98a,osad02a} 
interlayer transport regimes. The exact field dependence is determined 
by the Fermi surface geometry. 
Since the latter is generally anisotropic in the plane of the 
layers, the magnetoresistance strongly depends on the azimuthal 
orientation of the field \cite{lebe97,osad96,lee98b}. For the coherent 
regime, the theoretical predictions are in a good agreement with 
our results on the "clean" sample as well as with numerous other 
experiments \cite{kart04a}. This is, however, not the case for the 
weakly incoherent regime, as follows from the data on the "dirty" 
sample. The fact that its resistance is insensitive to the in-plane 
field is clearly in conflict with the existing theories 
\cite{kenz98a,osad02a}. 

An important point is that the anomalous behavior of 
the "dirty" sample  is observed in both the zero- and 
high-pressure states of $\alpha$-(BEDT-TTF)$_2$KHg(SCN)$_4$, 
characterized by different Fermi surface geometries. 
Moreover, a similar broad dip centered 
at $\theta = 90^{\circ}$ was found in the angle-dependent 
magnetoresistance of other highly anisotropic  
materials: the purely quasi-1D compound (TMTSF)$_2$PF$_6$ 
\cite{dann95a,kang04}, purely quasi-2D artificial GaAs/AlGaAs 
superlattice \cite{kura03}, and probably the 
most anisotropic of known organic conductors 
$\beta^{\prime\prime}$-(BEDT-TTF)$_2$SF$_5$CH$_2$CF$_2$SO$_3$, 
combining open and cylindrical Fermi surfaces \cite{wosn02}. 
Kuraguchi {\it et al.} \cite{kura03} already noted 
that a change in the interlayer transfer 
integral leads to a radical change in the magnetoresistance 
anisotropy although their data was not sufficient to 
establish the independence of the inplane field component. 

In conclusion, our data on the angle-dependent interlayer 
magnetoresistance of $\alpha$-(BEDT-TTF)$_2$KHg(SCN)$_4$ 
reveals a dramatic sample dependence which is most likely 
caused by the crossover between the coherent and weakly 
incoherent interlayer transport regimes. In the coherent 
regime the magnetoresistance is highly sensitive to both 
the polar and azimuthal orientations of the applied 
magnetic field that can be understood in terms of the 
conventional anisotropic 3D Fermi liquid theory. 
By contrast, in the weakly incoherent case the 
nonoscillating magnetoresistance background does not  
depend on the azimuthal orientation, in fields 
strongly inclined towards the layers, and 
can be scaled by a function of only the out-of-plane 
field component. 
This anomalous behavior appears to be a general 
feature of the weakly incoherent magnetotransport, 
regardless of the inplane Fermi surface geometry. 
However, the mechanism responsible for it remains unclear, 
indicating that a considerable modification of the 
existing theory is necessary. 

We thank P. Chaikin, A. Lebed, and P. Grigoriev for 
valuable stimulating discussions. The work was 
supported by the INTAS grant 01-0791, and the 
DFG-RFBR grant 436 RUS 113/592.

\end{document}